\documentclass[twocolumn,showpacs,preprintnumbers,amsmath,amssymb,prl,superscriptaddress]{revtex4}

\pdfoutput=1
\usepackage{graphicx}% Include figure files
\usepackage{dcolumn}% Align table columns on decimal point
\usepackage{bm}% bold math
\usepackage{booktabs}

\def\adot{\dot{\alpha}/\alpha}
\newcommand{\sss}{\scriptscriptstyle}

\makeatletter
\newcommand{\thickhline}{%
    \noalign {\ifnum 0=`}\fi \hrule height 1pt
    \futurelet \reserved@a \@xhline
}
\makeatother
\begin{document}

%\preprint{APS/123-QED}

\title{New limits on variation of the fine-structure constant using atomic dysprosium}

\author{N. Leefer}
 \affiliation{Department of Physics, University of California at
Berkeley, Berkeley, California 94720-7300, USA}
\author{C. T. M. Weber}
 \affiliation{Technische Universit\"{a}t Berlin, Berlin, Germany}
\author{A. Cing\"{o}z}
 \affiliation{AOSense, Sunnyvale, California 94085-2909, USA}
\author{J. R. Torgerson}
 \affiliation{Quantel USA, Bozeman, Montana 59715-1737, USA}
\author{D. Budker}
  \affiliation{Department of Physics, University of California at Berkeley, Berkeley, California 94720-7300, USA}
  \affiliation{Nuclear Science Division, Lawrence Berkeley National Laboratory, Berkeley, California 94720, USA}

\date{\today}
\begin{abstract}
We report on the spectroscopy of radio-frequency transitions between nearly-degenerate, opposite-parity excited states in atomic dysprosium (Dy). Theoretical calculations predict that these states are very sensitive to variation of the fine-structure constant, $\alpha$, owing to large relativistic corrections of opposite sign for the opposite-parity levels. The near degeneracy reduces the relative precision necessary to place constraints on variation of $\alpha$ competitive with results obtained from the best atomic clocks in the world. Additionally, the existence of several abundant isotopes of Dy allows isotopic comparisons that suppress common-mode systematic errors. The frequencies of the 754-MHz transition in $^{164}$Dy and 235-MHz transition in $^{162}$Dy were measured over the span of two years. Linear variation of $\alpha$ is found to be $\dot{\alpha}/\alpha = (-5.8\pm6.9)\times10^{-17}$~yr$^{-1}$, consistent with zero. The same data are used to constrain the dimensionless parameter $k_\alpha$, characterizing a possible coupling of $\alpha$ to a changing gravitational potential. We find that $k_\alpha = (-5.5\pm5.2)\times10^{-7}$, essentially consistent with zero and the best constraint to date.
\end{abstract}

%32.70.Jz: Line shapes, widths and shifts
%32.30.Bv: Radio-frequency, microwave, and infrared spectra
%06.20.Jr: Determination of fundamental constants
\maketitle
%\section{Introduction}

Variation of fundamental constants was first formulated by Dirac as the Large Numbers hypothesis~\cite{Dirac1937,Dirac1974}. The observation that dimensionless ratios of quantities such as the age of the universe to atomic time scales and the electromagnetic to gravitational force between a proton and electron were of the same order of magnitude, $\sim10^{40}$, led to the hypothesis that these ratios were functions of the age of the Universe. A consequence of this hypothesis is a gravitational constant, $G$, that scales inversely proportional to the age of the universe. Although modern experiments based on lunar ranging~\cite{Muller2007} have ruled out present-day variation of such magnitude, the variability of fundamental constants remains an active area of theoretical and experiment research. Any such variation would be a violation of the Einstein Equivalence Principle (EEP) and an indication of physics beyond General Relativity (GR) and the Standard Model (SM) of particle physics~\cite{Uzan2003,Uzan2011}.

Changing constants would manifest in a wide range of physical observables. The dimensionless electromagnetic-coupling constant, the fine-structure constant, $\alpha$, is of particular importance due to the implications of its variation on atomic clocks and time keeping. Any variation of $\alpha$ would lead to a change in the relative frequencies of co-located clocks even in the absence of external fields. This is forbidden by an assumption of EEP. In this letter we report new constraints on variation of $\alpha$ with respect to time and changing gravitational potential from a comparison of radio-frequency transitions in two isotopes of atomic dysprosium (Dy)~\cite{Budker1994,Nguyen2004}. These new results improve on our earlier constraints~\cite{Cingoz2007,Ferrell2007} by almost two orders of magnitude and are competitive with existing limits from other experiments~\cite{Guena2012,Peik2010,Tobar2010,Rosenband2008,Blatt2008,Fortier2007,Fischer2004,Peik2004}.

%Recent results from a comprehensive study of quasar absorption spectra indicates a possible spatial variation of the fine-structure constant, $\alpha$, at the level of $\Delta \alpha/\alpha \sim 10^{-6}$ Glyr$^{-1}$. An Earth-based laboratory moving through the galaxy would see this as a time variation at the level of $\dot{\alpha}/\alpha\,\sim\,10^{-19}$ yr$^{-1}$. The best laboratory experiment involving a comparison of Al$^+$ and Hg$^+$ optical transitions has constrained present day variation of $\alpha$ at the level of $|\dot{\alpha}/\alpha|\,\le\,4\times10^{-17}$ yr$^{-1}$. This introduction really sucks so finish it after the rest of the paper has been written

%\section{Method}

%\subsection{Clock comparisons}

The most stringent laboratory constraints on variation of fundamental constants come from clock-comparison experiments. We restrict our attention to clocks based on transitions in atoms and molecules. The ratio of any two such clock frequencies can be written as~\cite{Flambaum2004}

\begin{equation}\label{eqn1}
X = \frac{\nu_1}{\nu_2} = A\times\alpha^{K_\alpha}\mu_e^{K_{e}}\mu_q^{K_{q}},
\end{equation}
where $A$ is a dimensionless factor dependent on atomic structure, $\mu_e = m_e/m_p$ is the electron-proton mass ratio, and $\mu_q = m_q/\Lambda_{QCD}$ is ratio of the quark mass to QCD-mass scale. The dimensionless constants $\mu_e$ and $\mu_q$ are important for comparisons involving transitions with hyperfine structure~\cite{Guena2012,Fischer2004} or molecular transitions~\cite{Nijs2011}. The sensitivity coefficients, $K_{\alpha,e,q}$, depend on the particular frequency ratio under consideration. A summary of coefficients for various comparisons can be found in Table~\ref{table:1}.

\begin{figure}[t]
\includegraphics[width=8.4cm]{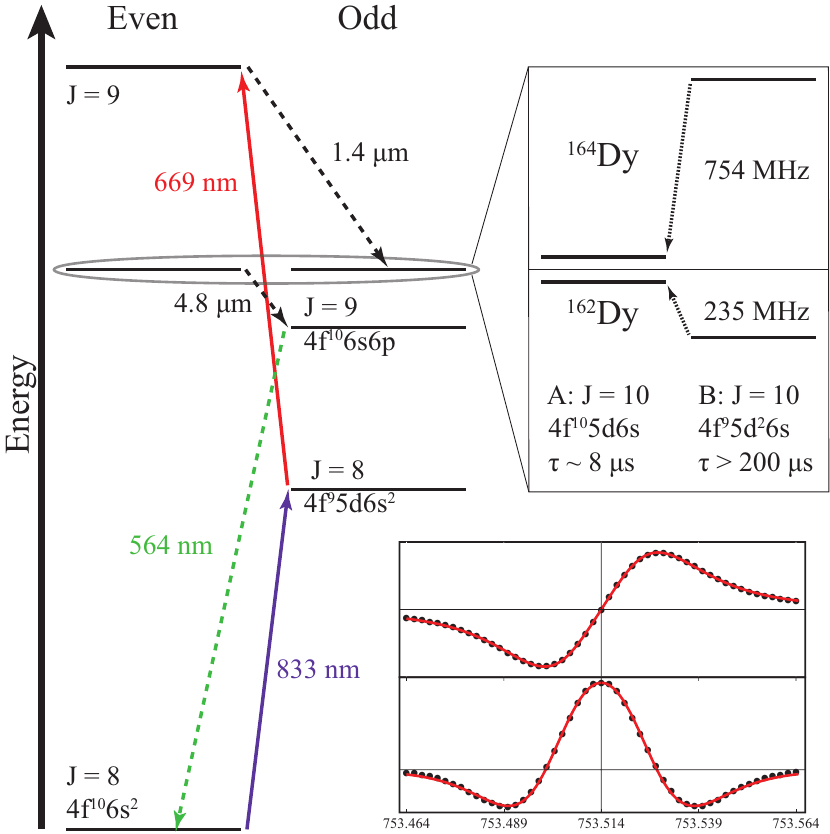}
\caption{\label{fig:1} Partial energy diagram for Dy showing states of interest. Preparation of atoms is accomplished via two laser excitations and a spontaneous decay with 30\% branching ratio into metastable state $B$. Atoms are excited from state $B$ to $A$ by a resonant, frequency-modulated rf electric field. State $A$ decays with lifetime $\sim8$ $\mu$s. A photomultipler tube and lock-in amplifier detect the 564-nm fluorescence. The bottom right inset shows typical lock-in signals for $^{164}$Dy at the 1st and 2nd harmonics of the modulation frequency.}
\end{figure}

In Dy we make use of an `accidental' degeneracy of energy levels to greatly relax the measurement precision necessary to place competitive limits on variation of $\alpha$. Large relativistic corrections to electron energies in Dy create an almost complete degeneracy of opposite-parity excited states, labeled $A$ and $B$ by convention (Fig.~\ref{fig:1}). This system has been the subject of investigations spanning over two decades, including an attempt to measure parity nonconservation~\cite{Nguyen1997,Dzuba2010}. Recently an analysis of the data from the present work has also been used to place stringent limits on violations of Lorentz symmetry and the Einstein Equivalence Principle~\cite{Hohensee2013}.

The energy difference corresponding, $\nu_{\sss BA} = (\epsilon_{\sss B} - \epsilon_{\sss A})/h$, is sensitive only to variation of $\alpha$. In practice, however, any measurement must have a standard `ruler' for comparison. The frequencies in our experiment are measured with respect to the stabilized oscillator of a cesium (Cs) beam standard, which introduces sensitivity to variation of both $\mu_e$ and $\mu_q$. Changes in the frequency ratio $\nu_{\sss BA}/\nu_{\sss\mathrm{Cs}}$ can be written

\begin{align}
\Delta \mathrm{ln}\frac{\nu_{\sss BA}}{\nu_{\sss\mathrm{Cs}}} = K_\alpha \frac{\Delta\alpha}{\alpha} + K_{\mu_e} \frac{\Delta\mu_e}{\mu_e} + K_{\mu_q} \frac{\Delta\mu_q}{\mu_q}.
\end{align}
As shown in Table~\ref{table:1}, the Dy/Cs frequency comparison is over six orders of magnitude more sensitive to variation of $\alpha$ than to variation of $\mu_e$ and $\mu_q$. This is a \textit{relative} enhancement of sensitivity, rather than an absolute enhancement, owing to the near degeneracy of levels $A$ and $B$ . At our present level of measurement precision, variation of $\mu_e$ or $\mu_q$ would only be observable at levels orders of magnitude larger than stringent constraints placed by other experiments~\cite{Guena2012}. Thus our experiment is effectively sensitive only to variation of $\alpha$,

\begin{equation}
\Delta \mathrm{ln}\frac{\nu_{\sss BA}}{\nu_{\sss\mathrm{Cs}}} = \frac{\Delta\nu_{\sss BA}}{\nu_{\sss BA}} - \frac{\Delta\nu_{\sss \mathrm{Cs}}}{\nu_{\sss \mathrm{Cs}}} \approx K_\alpha \frac{\Delta\alpha}{\alpha}.
\end{equation}
Instability of the Cs reference, a $>30$~yr old HP5061A, presents another source of concern for measurements spanning several years. A separate comparison between the Cs reference and a GPS stabilized Rb oscillator (Symmetricom TS2700) is performed during all data collection as a check against this. The fractional instability of the Cs reference, as compared to Rb reference, has been measured to be $<10^{-12}$~yr$^{-1}$, well below our dominant measurement errors. The influence of Cs-reference instability is ignored, and the magnitude of the frequency $|\nu_{\sss BA}|$ is assumed to vary with $\alpha$ according to~\cite{Dzuba2008}

\begin{equation}\label{eq:3}
\Delta |\nu_{\sss BA}| \approx \pm (2 \times10^{15}\,\mathrm{Hz})\,\Delta\alpha/\alpha,
\end{equation} 
where the sign is negative for $\nu_{\sss BA} > 0$ and positive for $\nu_{\sss BA}<0$. The present work is based on measurements of the $\nu_{\sss BA} \approx 753.5$~MHz and $\nu_{\sss BA} \approx  -234.7$~MHz transitions in $^{164}$Dy and $^{162}$Dy (see Fig.~\ref{fig:1}). Comparing isotopes with sensitivities of opposite sign allows for the cancellation of common systematic errors that might otherwise mimic variation of $\alpha$ in a single isotope.

\begin{table}[t]
  \centering
  \begin{tabular*}{0.47\textwidth}{@{\extracolsep{\fill}} l  r  r r r}
  \thickhline\noalign{\vspace{0.15mm}}\thickhline
  ratio & $K_\alpha$ & $K_e$ & $K_q$ & ref.\\
  \thickhline\noalign{\smallskip}
$^{164,162}$Dy/Cs & $(-2.6,+8.5)\times10^6$  & $-1$ & $-0.002$ &[this work]  \\
Rb/Cs & $-0.49$ & $0$ & $-0.021$ & \cite{Guena2012}\\
Yb$^+$/Cs & $-1.83$ & $-1$ & $-0.002$ & \cite{Peik2010}\\
CSO/Cs & 3 & $-1$ & $0.1$ & \cite{Tobar2010}\\
Hg$^+$/Al$^+$ &  $-2.95$ & $0$ & $0$ & \cite{Rosenband2008}\\
Sr/Cs & $-2.77$ & $-1$ & $-0.002$ & \cite{Blatt2008}\\
H(1S-2S)/Cs& $-2.83$ & $-1$ & $-0.002$ & \cite{Fischer2004}\\

  \thickhline\noalign{\vspace{0.15mm}}\thickhline
\end{tabular*}
\caption{Sensitivity coefficients for several clock comparisons. CSO refers to crystal-sapphire oscillator. The large sensitivity of the Dy transition frequency to variation of $\alpha$ is a relative enhancement due to the near degeneracy of the electronic states involved in the transition. Column references are for experimental details. Calculations of sensitivity coefficients can be found in Refs.~\cite{Dzuba2008,Dinh2009}.}
\label{table:1}
\end{table}

\begin{figure}[t]
\includegraphics[width=8.4cm]{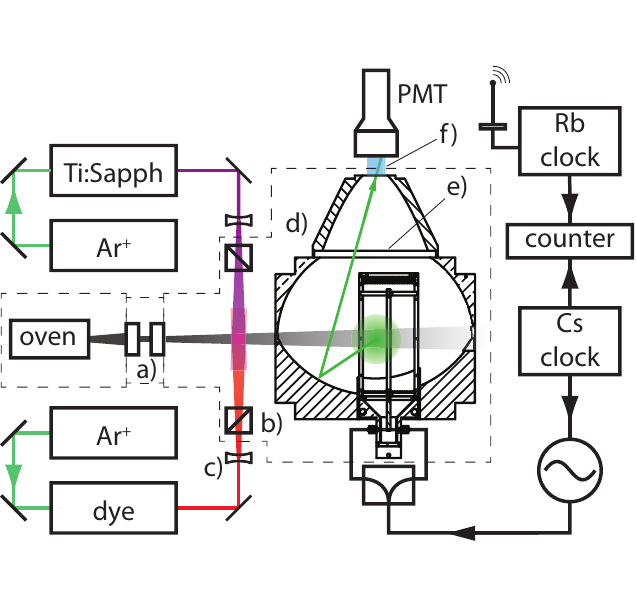}
\caption{\label{fig:2} Schematic of the experimental set-up. Argon-ion lasers pump a dye laser producing 669-nm light and a Ti:sapphire laser producing 833-nm light. Components in vacuum are within the dashed boundaries. a) Skimmers collimate the atomic beam, and double as conductance chokes for differential pumping between the oven chamber and interaction chamber. b) In-vacuum linear polarizers are the last optical element for the laser light before interacting with Dy atoms. c) Lenses diverge the laser light to match the atomic beam divergence. d) Polished aluminum mirrors guide fluorescence to a photomultiplier tube. e) An interference filter wih 564-nm peak transmission suppresses stray laser and oven light. f) A glass pipe guides fluorescent light to a PMT for detection.}
\end{figure}

The spectroscopy is performed on a thermal beam of Dy atoms, produced in an oven heated to $\sim1400$~K inside a vacuum chamber with residual gas pressure of $\sim10^{-7}$~torr. After two collimators/conductance chokes the atoms enter the interaction chamber where the residual gas pressure is $\sim10^{-9}$~torr. The atoms undergo laser excitations at 833 nm and 669 nm, employing an adiabatic-passage technique~\cite{Nguyen2000}, followed by a spontaneous decay at 1.4 $\mu$m with 30\% branching ratio to state $B$. Narrow-band lasers provide high-fidelity isotope selection. Upon excitation to state $B$ atoms then enter the interaction region, where excitation from $B$ to $A$ occurs via a frequency-modulated electric field. Atoms spontaneously decay from state $A$ via two steps to the ground state. Fluorescence at 564 nm is directed by a polished-aluminum light-collection system ($\sim4\%$ overall efficiency) into a glass pipe, detected by a photomultiplier tube (PMT), and sent to a lock-in amplifier for processing. Figure~\ref{fig:2} shows a simplified diagram of the experiment.

%\subsection{Systematics}

The apparatus has been designed to minize the systematic uncertainties presented in Table~\ref{table:2}. In our previous result~\cite{Cingoz2007} the sensitivity was limited by the collisional pertubation of energy levels by background gases~\cite{Cingoz2005}, poor suppression of Zeeman shifts owing to imperfections in laser-light polarization, and systematic shifts related to inhomogeneity of the rf field. In the new apparatus, the high-vacuum system reduces collisional shifts to below the 1~mHz level. The electric-field region has been designed to ensure field homogeneity across the range of operating frequencies. Doppler shifts are suppressed by creating an rf-standing wave in the interaction region in addition to orienting the k-vector of any residual traveling wave perpendicular to the atomic-beam propagation axis. Two layers of magnetic shielding limit background magnetic fields to below $500\,\mu$G in all directions and three-axis magnetic field coils allow residual fields to be canceled out.

%A different quadrupole moment of the electron charge distribution in states $A$ and $B$ would give rise to systematic frequency shifts in the presence of an electric-field gradient. Although not measured, this effect is estimated by assuming a typical quadrupole moment of 1 $e\,a_0^2$ an electric field gradient of $0.075$~V cm$^{-2}$, calculated from finite-element modeling of the electric field plates. 

The dominant systematic is an electronic offset in the acquisition electronics, which may create a shift in the zero-crossing of the first-harmonic signal and apparent shift of the transition frequency. Sensitivity to electronic offsets is amplified by the relatively large transit-broadened linewidth of the transition, $\gamma \sim 2 \pi \times40\,\mathrm{ kHz}$. We measure these offsets by varying the PMT gain, in order to change the useful signal size while leaving electronic noise unchanged. This idea is based on the offset compensation scheme presented in Ref.~\cite{Mueller2003}, but currently only constrains electronic offsets at the level of 500 nV. The importance of this effect depends on the absolute signal size and is reflected in the range of uncertainties in Table~\ref{table:2}.

\begin{table}[t]
  \centering
  \begin{tabular*}{0.47\textwidth}{@{\extracolsep{\fill}}l r r}
  \thickhline\noalign{\vspace{0.15mm}}\thickhline\noalign{\smallskip}
  systematic & stability (mHz) & $|\dot{\alpha}/\alpha|$ ($10^{-17}\,yr^{-1}$) \\
  \thickhline\noalign{\smallskip}
  electronic offsets & 200-470 & 10-23.5 \\
  BBR/temperature & 66 & 3.3 \\
  Zeeman shift & 50 & 2.5 \\
  ac-Stark shift & 32 & 1.6 \\
  res. amp. mod. & 20 & 0.5 \\
 % Millman effect & $<10$ & 0.4 \\
  dc-Stark shift & $<1$ & $<0.04$ \\
  collisional shift & $<1$ & $<0.04$ \\
  quadrupole shift & $<1$ & $<0.04$ \\
  clock stability & $<1$ &$<0.04$\\
  \thickhline\noalign{\smallskip}
  \textbf{Total} & \textbf{220 - 480} & \textbf{11 - 24}\\
  \thickhline\noalign{\vspace{0.15mm}}\thickhline
\end{tabular*}
\caption{Current levels of known systematics. The total systematic uncertainty is obtained by adding in quadrature. The corresponding uncertainties for $|\dot{\alpha}/\alpha|$ assume two measurements separated by one year.}
\label{table:2}
\end{table}

The ac-Stark shift in a two-level system is approximately zero for a resonant electric field, with a negligible contribution expected from what is known as the Bloch-Siegert shift~\cite{Nguyen2004}. Strongly coupled off-resonant levels may lead to large shifts correlated with rf-power. A measurement of the off-resonant contributions to the dynamic polarizabilities in $^{164}$Dy and $^{162}$Dy found $\delta\nu \simeq 70\,E^2$~mHz, where $E^2$ is mean-squared field value. Typical values of $E^2$ are 4.5 (V/cm)$^2$, corresponding to a stability of 3 mHz/\% change in rf power. The uncertainty associated with this systematic is conservatively estimated from an assumption of 10\% control over the rf power in the interaction region.

Additional Stark related systematics are the dc-Stark effect and blackbody radiation (BBR) induced Stark shifts~\cite{Safronova2010a}. Charged particles in the atomic beam can cause charge accumulation on the electric field plates and produce DC fields. An electrode biased at 500 V is used to sweep charged particles out of the atomic beam, and the DC field is periodically measured via Zeeman-crossing spectroscopy~\cite{Budker1994,Nguyen1997}. These are consistently found to be at the level of 10 mV/cm. The temperature dependence of the transition frequencies has been measured near room temperature to be $+29(4)$~mHz/K and $-34(4)$~mHz/K for $^{164}$Dy and $^{162}$Dy, respectively. The isotopic dependence of the sign is consistent with BBR induced Stark shifts, but the attribution of these shifts to BBR is preliminary~\cite{Weber2013}. Currently, the 2 K temperature stability of the interaction region is used to estimate the systematic uncertainty due to this effect. 

Suppression of systematics related to Zeeman shifts is accomplished by performing spectroscopy with the Zeeman structure unresolved. Linear polarizers are located in vacuum and are the last optical elements for the 833-nm and 669-nm laser light, ensuring symmetric population of the $\pm M$ magnetic sublevels of state $B$. A magnetic field then leads to a broadening of the unresolved line, but no shift.  A measured residual Zeeman shift of $\sim2.5$ Hz/mG represents a suppression of $\sim1000$ from the sensitivity of the $m=10$ sublevel. The magnetic field stability along the quantization axis, chosen to coincide with the rf field, is at the level of $20\,\mu$G.  We note that the magnetic field insensitive $m_{\sss B} = 0 \rightarrow m_{\sss A} =0$ transition is forbidden between levels $A$ and $B$ where $\Delta J = 0$.

Residual amplitude modulation refers to a power imbalance of the carrier sidebands in the frequency-modulated spectrum of the electric field. Such an imbalance distorts the atomic lineshape and creates an apparent frequency shift. Poor impedance matching and termination of the rf transmission line made this a dominant systematic in early data at the 1 Hz level. Measuring the transition frequency with the in-phase and quadrature channels of the lock-in amplifier allows the size and stability of RAM to be measured directly with the atoms~\cite{LeeferThesis}. This protocol was implemented beginning May 2011. In August 2011, custom narrow-band radio-frequency circulators (DPV CO) were acquired to suppress transmission-line etalons, reducing RAM to the level of $\sim10$~ppm. The frequency shift introduced by this modification was measured and a correction applied to earlier data.

%An attempt to measure this directly by varying the electric field amplitude yielded $\delta\nu = 0.03(3)$ Hz (V/cm)$^{-2}$ 
%\subsection{Variation of $\alpha$}

\begin{figure}[t]
\includegraphics[width=8.5cm]{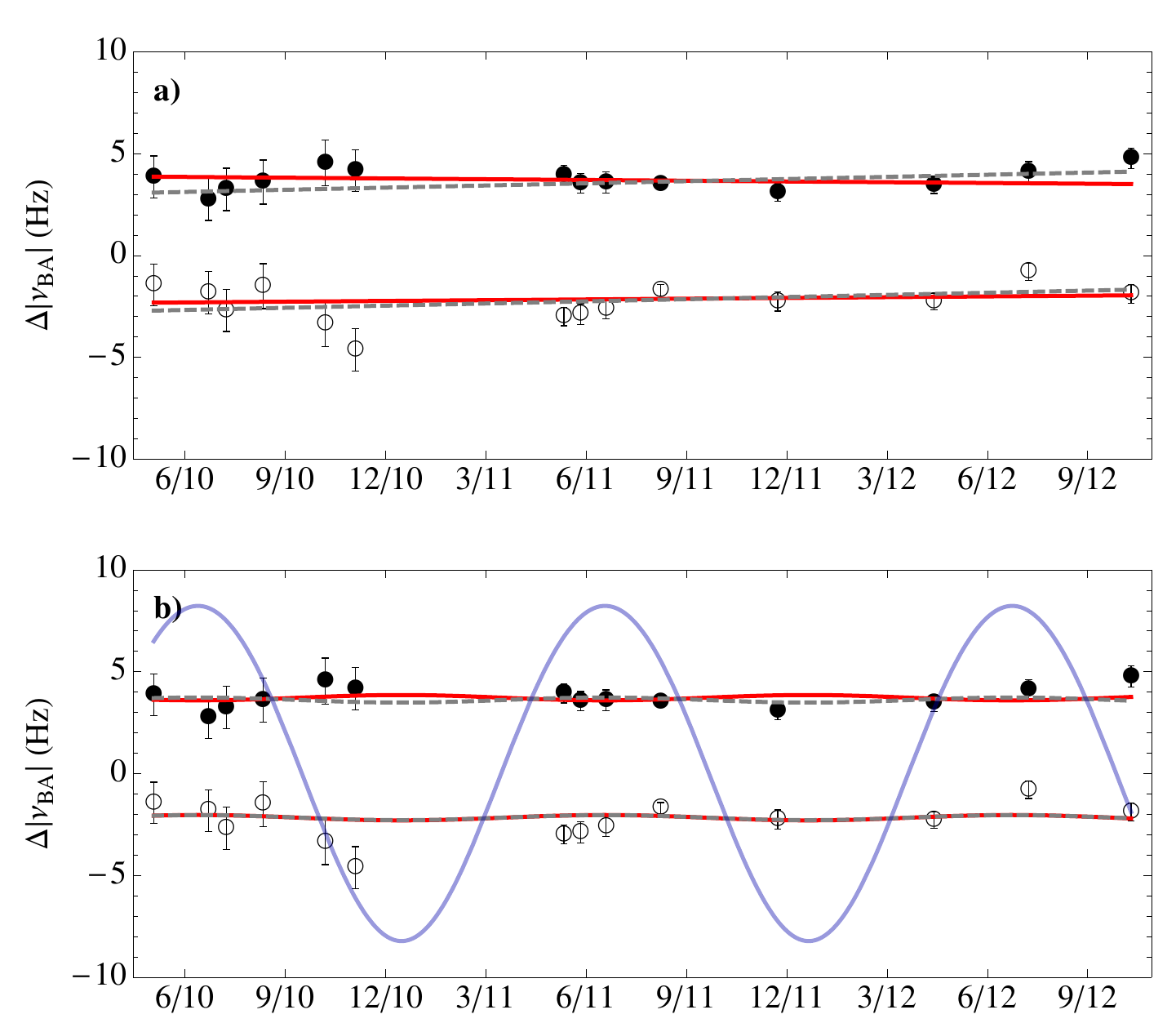}
\caption{\label{fig:3} Changes in the transition frequencies for $^{162}$Dy (filled circles) and $^{164}$Dy (empty circles) over the span of two years. The frequencies for $^{162}$Dy and $^{164}$Dy are displayed with respect to 234,661,102 Hz and 753,513,708 Hz, respectively. a) The data are fit by linear functions with equal magnitude slopes of opposite sign (solid) and same sign (dashed). b) The data are fit by cosine functions with equal amplitudes but 180$^\circ$ phase difference (solid) and 0$^\circ$ phase difference (dashed). The variation of the dimensionless gravitational potential, scaled in relative units by $5\times10^{10}$, is shown by the light solid line.}
\end{figure}

The transition frequencies $\nu_{164}$ and $\nu_{162}$ measured over the span of two years are shown in Fig.~\ref{fig:3}. The reduced uncertainties beginning in May 2011 are primarily due to the characterization and eventual suppression of RAM. To constrain a linear variation of $\alpha$ in time, a global linear least-squares fit is performed, in which the two isotopes' data are fit by independent offsets and equal magnitude slopes of opposite sign. The best-fit slope of $-0.12 \pm0.14$~Hz/yr corresponds to the result

\begin{equation}
\adot = (-5.8\pm6.9)\times10^{-17}\,\mathrm{ yr}^{-1},
\end{equation}
which is consistent with zero within 1 standard deviation. This result approaches within a factor of 3 the level obtained with the best optical clocks in the world~\cite{Rosenband2008}, and is limited by systematics. The contribution of statistical uncertainties is at the level of $\adot\sim1.7\times10^{-17}$~yr$^{-1}$. The data are also fit by equal slopes of the same sign, which is sensitive to common mode systematics, but not variation of $\alpha$. This fit gives a slope of $0.41\pm0.14$~Hz/yr. The 3-sigma, non-zero drift at the level of $\sim0.5$~Hz could be explained by a drifting electronic offset which, as a technical systematic, is expected to be the same sign for both isotopes.

Our data can also be used to constrain violations of local position invariance, assuming a model where fundamental constants are influenced by light scalar fields that scale linearly with changes in the local gravitational potential~\cite{Flambaum2007}. We can express this as $\Delta\alpha/\alpha = k_\alpha \Delta U/c^2$, where $\Delta U/c^2$ is a change in the dimensionless gravitational potential. The ellipticity of the Earth's orbit provides semi-annual changes in the laboratory gravitational potential, $\Delta U/c^2 = \pm1.65\times10^{-10}$, at the aphelion and perihelion of Earth's orbit for plus and minus signs, respectively. To constrain $k_\alpha$ the data are again fit by global linear least-squares to cosine functions with equal amplitudes but $180^\circ$ phase difference. The period is equal to one solar year and zero phase is fixed at Earth's perihelion on Jan. 3, 2010. The best-fit amplitude of oscillation is found to be $0.18\pm0.17$~Hz, providing the best constraint to date on the dimensionless coupling parameter~\cite{Guena2012}
\begin{equation}
k_\alpha =(-5.5\pm5.2)\times10^{-7},
\end{equation}
which is also consistent with zero at $\sim$1 standard deviation. The sensitivity is again limited by systematics. The statistical contribution to the uncertainty is at the level of $k_\alpha \sim 1.2\times10^{-7}$.  A global fit to the two isotopes' data with $0^\circ$ phase difference, sensitive to common mode systematics, has an amplitude of $-0.17\pm0.17$~Hz. The data and best fits are shown in Figure~\ref{fig:3}

%\section{Conclusion}

We have presented updated constraints on variation of $\alpha$ that represent almost two orders of magnitude improvement over previous results, with the present level of sensitivity still limited by systematic effects. While more stringent control of these systematics, particularly electronic offsets, presents a clear avenue to achieving the ultimate practical statistical limit of $10^{-18}$ calculated in~\cite{Nguyen2004}, recent astrophysical evidence~\cite{Webb2011} for spatial variation of $\alpha$ suggests an observable variation of $\alpha$ in the laboratory at the level of $10^{-19}$~\cite{Berengut2012p1}. A new generation of experiments based on the spectroscopy of optical nuclear transitions~\cite{Berengut2012p2} or optical transitions in highly-charged ions~\cite{Berengut2012p3} will be necessary to observe this effect.

The authors are grateful to  A. Lapierre, A.-T. Nguyen, S. K. Lamoreaux, Uttam Paudel, and V. V. Yacschuk for crucial contributions to the earlier stages of the experiment, V. A. Dzuba and V. V. Flambaum for supporting atomic structure calculations and discussions, and D. English and M. Hohensee for many invaluable discussions. D.B. acknowledges support from the Miller Institute for Basic Research in Science. This research has been supported by the National Science Foundation and Foundational Questions Institute.

\bibliography{prl2012refs}

\end{document}